\documentclass[12pt]{spieman}  
\usepackage{amsmath,amsfonts,amssymb}
\usepackage{graphicx}
\usepackage{setspace}
\usepackage{tocloft}
\usepackage{lineno}
\usepackage{upgreek}
\usepackage{color, soul}

\title{Alignment and optical verification of DESHIMA 2.0 at ASTE}

\author[a,*]{Arend Moerman}
\author[a,b]{Kenichi Karatsu}
\author[a,b,c]{Jochem J. A. Baselmans}
\author[a,b]{Shahab O. Dabironezare}
\author[d]{Shinji Fujita}
\author[b]{Robert Huiting}
\author[e,f]{Kotaro Kohno}
\author[e]{Yuri Nishimura}
\author[g]{Fenno Steenvoorde}
\author[h]{Tatsuya Takekoshi}
\author[i]{Yoichi Tamura}
\author[h]{Akio Taniguchi}
\author[j]{Stephen J. C. Yates}
\author[k,l]{Bernhard R. Brandl}
\author[a]{Akira Endo}
\affil[a]{Faculty of Electrical Engineering, Mathematics and Computer Science, Delft University of Technology, Mekelweg 4, 2628 CD Delft, The Netherlands}
\affil[b]{SRON—Netherlands Institute for Space Research, Niels Bohrweg 4, 2333 CA Leiden, The Netherlands}
\affil[c]{Physikalisches Institut, Universität zu Köln, Zülpicher Straße 77, 50937 Cologne, Germany}
\affil[d]{The Institute of Statistical Mathematics, 10-3 Midori-cho, Tachikawa, Tokyo 190-8562, Japan}
\affil[e]{Institute of Astronomy, Graduate School of Science, The University of Tokyo, 2-21-1 Osawa, Mitaka, Tokyo 181-0015, Japan}
\affil[f]{Research Center for the Early Universe, Graduate School of Science, The University of Tokyo, 7-3-1 Hongo, Bunkyo-ku, Tokyo 113-0033, Japan}
\affil[g]{DEMO: Electronic and Mechanical Support Division, Delft University of Technology, Mekelweg 4, 2628 CD Delft, The Netherlands}
\affil[h]{Kitami Institute of Technology, 165 Koen-cho, Kitami, Hokkaido 090-8507, Japan}
\affil[i]{Department of Physics, Graduate School of Science, Nagoya University, Furo-cho, Chikusa-ku, Nagoya, Aichi 464-8602, Japan}
\affil[j]{SRON—Netherlands Institute for Space Research, Landleven 12, 9747 AD Groningen, The Netherlands}
\affil[k]{Leiden Observatory, Leiden University, PO Box 9513, 2300 RA Leiden, The Netherlands}
\affil[l]{Faculty of Aerospace Engineering, Delft University of Technology, Kluyverweg 1, 2629 HS Delft, The Netherlands}

\cftpagenumbersoff{figure}
\cftpagenumbersoff{table} 
\begin{document} 
\maketitle

\begin{abstract}
We developed, characterized, and verified an alignment procedure for the DESHIMA 2.0 instrument, an ultra wide-band spectrometer operating between 200--400 GHz, at the ASTE telescope. To this end, we mounted the warm optics, consisting of a modified Dragonian dual reflector system, on a motor controlled hexapod. Crucial in the alignment procedure is our sky chopper, which allows fast beam switching. It has a small entrance and exit aperture coupling to (cold) sky, which creates a measurable signal with respect to the warm cabin environment. By scanning the instrument beam across the entrance aperture of the sky chopper using the hexapod, we found the hexapod configuration that produced the lowest signal on our detectors, implying the beam is coupled fully to cold sky and not the warm cabin. We first characterized the alignment procedure in the laboratory, where we used a vat containing liquid nitrogen as the cold source behind the sky chopper. Then, we applied the alignment procedure to DESHIMA 2.0 at ASTE. We found that the alignment procedure significantly improved the aperture efficiency compared to previously reported values of the aperture efficiency of DESHIMA at ASTE, which indicates the veracity of the alignment procedure. 
\end{abstract}


{\noindent \footnotesize\textbf{*}Corresponding author,  \linkable{A.Moerman-1@tudelft.nl} }


\section{Introduction}
\label{sect:intro}  
The Deep Spectroscopic High-redshift Mapper (DESHIMA) is an ultra wide-band (UWB) submillimeter (submm) integrated superconducting spectrometer (ISS) using MKIDs~\cite{Day2003}. The primary design (DESHIMA 1.0~\cite{Endo_2019B}) has instantaneously covered 332--377~GHz by 49 frequency channels, and the demonstration instrument has succesfully seen first light in 2017~\cite{Endo_2019A} at the 10-meter Atacama Submillimeter Telescope Experiment (ASTE)~\cite{Ezawa_2004} telescope in Chile. DESHIMA 2.0 is a successor to DESHIMA 1.0 with the upgrades including an increase of the bandwidth (200--400~GHz), the number of frequency channels (347 channels), the optical efficiency, and the addition of a fast sky-position chopper for atmospheric subtraction, among others~\cite{Taniguchi_2022}. The science cases range from multi-line spectroscopy of dusty star-forming galaxies in order to calculate spectroscopic redshifts and study their physical conditions~\cite{Rybak2022}, broad-band continuum observations of asymptotic giant branch stars, to galaxy cluster diagnostics through the Sunyaev-Zel'dovich effect~\cite{Sunyaev1970,Sunyaev1980}.

The DESHIMA 2.0 design is based on a superconducting on-chip filterbank~\cite{PascualLaguna2021,Thoen2022}, fed by a leaky-lens antenna~\cite{Neto2010a, Neto2010b,Hahnle2020}. The filterbank is connected to an NbTiN-Al hybrid MKID~\cite{Janssen2013} array, which is read-out using frequency-domain multiplexing~\cite{vanRantwijk2016}. The integrated nature of the DESHIMA architecture and the multiplexed readout makes it inherently scalable and as such, it plays an important role in future integral field unit (IFU) designs~\cite{Jovanovic2023}, such as the Terahertz Integral Field Unit with Universal Nanotechnology (TIFUUN) instrument which is currently in its early stages of development~\cite{Mavropoulou2023}. The cryostat housing DESHIMA 2.0 is an adiabatic demagnetization refrigerator (ADR), where it operates at a temperature of 120 mK.

The submillimeter beam out of the vacuum window of DESHIMA can be misaligned from the design by as much as $\sim$4 degrees~\cite{Endo_2019B}, which can lead to a significant degradation of the aperture efficiency when mounted on the ASTE telescope without sufficient correction~\cite{Endo_2019B}. Furthermore, because the cold optics are partially disassembled during transportation, there is no guarantee that the misalignment will be persistent from the laboratory to the telescope, requiring measurement of the misalignment in the telescope cabin. Fortunately, DESHIMA 2.0 employs a modified Dragonian dual reflector~\cite{Dragone_1978} at room temperature for magnification and coupling to the ASTE subreflector, and the positions of these mirrors can be adjusted to correct for potential misalignment. 

Such a correction for misalignment by one or more room-temperature optical components are common in mm/submm astronomy, though the exact methodology can vary depending on the system. A common method employs an optical laser to align the optics at the telescope. Examples include CONCERTO at APEX~\cite{Ade2020,Catalano2022} and NIKA2 at IRAM~\cite{Adam2018}. However, this is an indirect method that assumes the submillimeter beam to be aligned to the laser ray, which is not the case for DESHIMA 2.0. The SOFIA instrument has used radiometric alignment by introducing an additional component that intercepts the beam~\cite{Graf2016}. An alternative approach is to measure the mechanical positions of the optical components by means of a Faro measurement arm and correct for any misalignment~\cite{Reyes}, though the position of the cryogenic mirrors cannot be measured.

Recently, Moerman et al.~\cite{Moerman2024} have demonstrated a method for aligning the DESHIMA 2.0 instrument in the laboratory. They measured the complex (phase and amplitude) submm beam to extract misalignment information~\cite{MingTangChen,Tong2003} and used this, in combination with a ray-tracing algorithm, to align the instrument using a modified Dragonian dual reflector and a mechanical hexapod. They argued that this method can correct for lateral shifts and tilt, but that a separate subreflector alignment should be performed to correct for the vertical shift, or defocus. However, applying this method at ASTE would require installing a coherent submm source and an XY-scanning stage in the small ASTE receiver cabin. Moreover, the presence of the aforementioned sky-position chopper significantly complicates the application of the alignment method of Moerman et al.~\cite{Moerman2024}.

In this paper we describe an alignment method that directly uses the instrument beam for alignment, without the need for quantification of misalignment using a visible light laser or a complex beam pattern measurement. The new method makes use of the temperature contrast between the room temperature entrance and exit apertures of the sky chopper and a cold background source, in combination with the modifed Dragonian dual reflector and mechanical hexapod also used by Moerman et al.~\cite{Moerman2024}. The main advantage of the proposed method is the lack of need for elaborate measurement setups, making it simpler to execute than the procedure in Moerman et al.~\cite{Moerman2024}. Instead, only a cold source, such as the atmosphere, is necessary. This makes the proposed method attractive for implementation at telescopes with limited receiver cabin space, such as the ASTE telescope. The method only has to be performed once after installation of the dual reflector, hexapod, and sky chopper. Also, the method can be executed fully remotely over an internet connection, eliminating the need to be physically present at the telescope site. This allows for remote checking of the alignment whenever it is desired.

The paper is structured as follows. We start by briefly discussing the optical path of DESHIMA 2.0 at ASTE in Sect.~\ref{subsect:opticalpath}. Particular focus here will be given to the sky chopper. In Sect.~\ref{subsect:genmethod} we discuss the general methodology behind the alignment method. Additionally, we discuss the alignment of the ASTE subreflector in Sect.~\ref{subsect:subrefalign}. In Sect.~\ref{subsect:expsetup} we show the laboratory setup used for characterisation of the alignment method. We also show the telescope setup at ASTE that was actually aligned using the proposed method. In Sect.~\ref{subsect:labres} We show the results from the laboratory characterisation and in Sect.~\ref{subsect:telres} we show results from the actual alignment at ASTE.

\section{Methods}
\subsection{Optical path at ASTE}
\label{subsect:opticalpath}
\begin{figure}[ht]
\centering
    \includegraphics[width=1\textwidth]{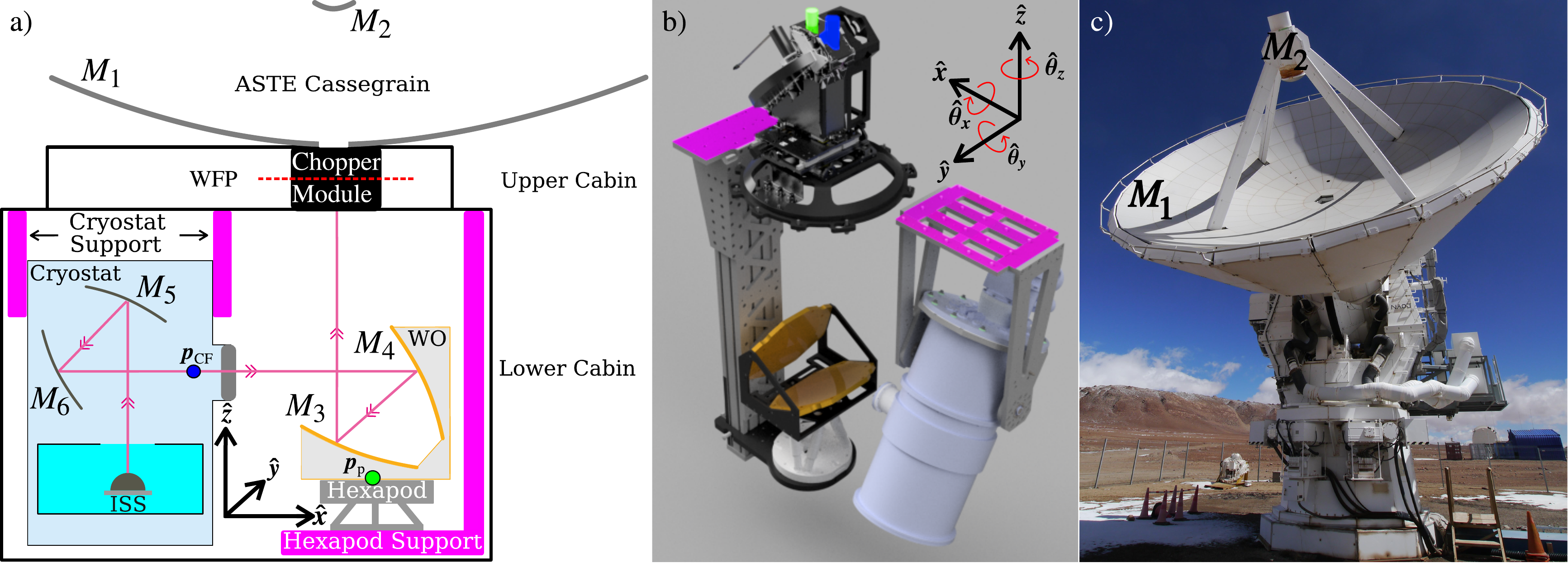}
    \caption{An overview of the optical setup of DESHIMA 2.0 at ASTE including the sky chopper. a) Sketch of the setup at the ASTE telescope. Illustrated are the integrated superconducting spectrometer (ISS), cryostat and cryostat optics $M_6$ and $M_5$, the cold focus $\pmb{p}_\mathrm{CF}$ (blue dot), the warm optics $M_4$ and $M_3$, the hexapod, support structures for the hexapod and cryostat, and the chopper module. The chopper module is located in the upper cabin of ASTE, the rest in the lower cabin. The warm focal plane, denoted WFP in the sketch, is illustrated with a red dashed line and runs through the sky chopper. The coordinate axes shown are used throughout the paper as the basis coordinates. The origin of the coordinate system is the cold focus, unless stated otherwise. The hexapod pivot $\pmb{p}_\mathrm{p}$, denoted by the green dot, is located in the center of the hexapod stage. The Cassegrain setup at ASTE, after the chopper module, consists of a hyperboloid subreflector $M_2$ and a paraboloid primary $M_1$. b) CAD render of the setup in panel a), showing the sky chopper and the two optical paths (in green and blue) coming out of the top of the chopper. We show the coordinate axes again to emphasise the orientation of the system. Also, we show (right-handed) rotation axes in this coordinate system. c) Photograph of the ASTE telescope in the Atacama desert, Pampa la Bola, Chile. Both $M_2$ and $M_1$ are illustrated in the figure. The lower cabin is visible below $M_1$. The hexagonal hole in the vertex of $M_1$ is the entrance of light from the sky into the upper cabin.}
    \label{fig:sketch_setup}
\end{figure}

We start by briefly describing relevant elements in the optical path of DESHIMA 2.0 at ASTE. A sketch of the setup and global coordinate system can be found in Fig.~\ref{fig:sketch_setup}a and a CAD render in Fig.~\ref{fig:sketch_setup}b. A photograph of the ASTE telescope can be found in Fig.~\ref{fig:sketch_setup}c.

 The first optical components seen from the DESHIMA leaky-lens antenna are two off-axis paraboloidal mirrors, $M_6$ and $M_5$, which are placed in a parabolic relay~\cite{Dabironezare} configuration. This compensates for the inherent aberration introduced by a single off-axis paraboloidal mirror~\cite{Murphy1987}. One focus of the parabolic relay overlaps the leaky-lens antenna, the other is located to the right of the relay in Fig.~\ref{fig:sketch_setup}a and denoted the cold focus (shown as $\pmb{p}_\mathrm{CF}$). Both $M_6$ and $M_5$ are located inside the cryostat, and are therefore referred to as the cold optics. The 30$^\circ$ rotation of the cryostat around the $x$-axis (see Fig.~\ref{fig:sketch_setup}b) is applied so that the cryostat is upright when the telescope is tilted to an elevation of 60$^\circ$. This ensures that the cryostat is mostly upright during regular observations.
 
 After the cold optics, the beam of the ISS encounters the warm optics. It consists of a hyperboloidal mirror $M_4$ and an ellipsoidal mirror $M_3$, which are placed in a modified Dragonian~\cite{Dragone_1978} setup and monolithically mounted onto a motor controlled hexapod. One focus of the warm optics overlaps the cold focus, the other is situated above the warm optics and is called the warm focus. The warm focus is enclosed within the sky chopper and lies in the warm focal plane (WFP). The WFP normal is oriented along the $z$-axis. 
 
\begin{figure}[t]
\centering
    \includegraphics[width=0.5\textwidth]{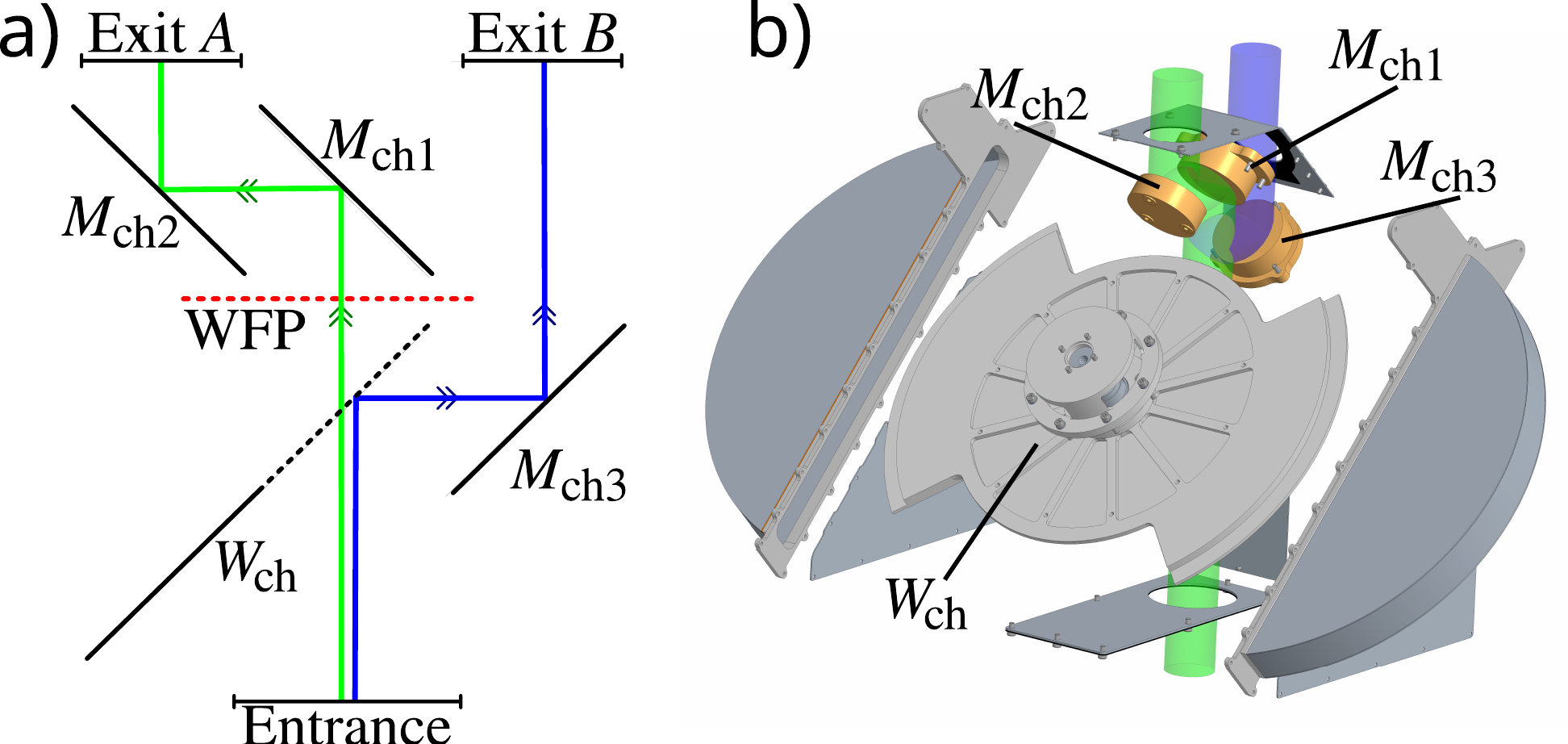}
    \caption{Overview and sketch of the sky chopper. a) Schematic sketch showing the location of the chopper wheel $W_\mathrm{ch}$ and the three planar mirrors $M_\mathrm{ch1}$, $M_\mathrm{ch2}$, and $M_\mathrm{ch3}$. On the bottom of the sketch we denote the entrance aperture. Both exit apertures $A$ and $B$ are also shown. Note also that the color coding of the beam through $A$ and $B$ matches the colors of the beams coming out of the sky chopper in Fig.~\ref{fig:sketch_setup}b. The warm focal plane, denoted WFP, runs through the middle of the sky chopper and is denoted by the horizontal red dashed line. b) Exploded view of a CAD render of the chopper module, showcasing the contents of the sky chopper. The illustrated green and blue tube do not represent the actual shape of the DESHIMA 2.0 beam, but are for illustrative purposes only.}
    \label{fig:chopper}
\end{figure}

 The sky chopper is a device containing three planar mirrors, $M_\mathrm{ch1}$, $M_\mathrm{ch2}$, $M_\mathrm{ch3}$, and a slotted, reflective wheel, which we will denote $W_\mathrm{ch}$. The beam enters the sky chopper through the entrance aperture, after which it encounters $W_\mathrm{ch}$. As $W_\mathrm{ch}$ rotates, half of the time the beam passes through the slots in the wheel, onto $M_\mathrm{ch1}$ and $M_\mathrm{ch2}$, leaving the sky chopper through exit aperture $A$. We will denote the beam coming out of exit aperture $A$ beam $A$. If $W_\mathrm{ch}$ is closed, the beam reflects off $W_\mathrm{ch}$, onto $M_\mathrm{ch3}$, and out through exit aperture $B$. This beam we denote beam $B$. For a detailed overview of the sky chopper and its contents, see Fig.~\ref{fig:chopper}. The entrance and exit apertures fulfill a crucial role in the alignment procedure by constraining the beam location and tilt as it passes through the sky chopper, essentially coupling the beam to the warm sky chopper aperture/interior and cabin environment if the alignment is poor, and coupling it to the cold sky if the alignment is good. In Sect.~\ref{subsect:genmethod} we will explain this in more detail. 
 
 After exiting the chopper module, both beams encounter $M_2$, the secondary (hyperboloidal) reflector of the ASTE Cassegrain. After $M_2$, the beams encounter $M_1$, the paraboloidal primary of the ASTE Cassegrain. Beam $A$ ($B$) illuminates $M_5$ 50 ($-$50) mm off-centre along the $x$-axis. This causes beam $A$ ($B$) to have a 117 ($-$117) arcsecond pointing offset along the azimuthal axis with respect to broadside, which is fixed by design.

\begin{figure*}[h]
\centering
    \includegraphics[width=0.8\textwidth]{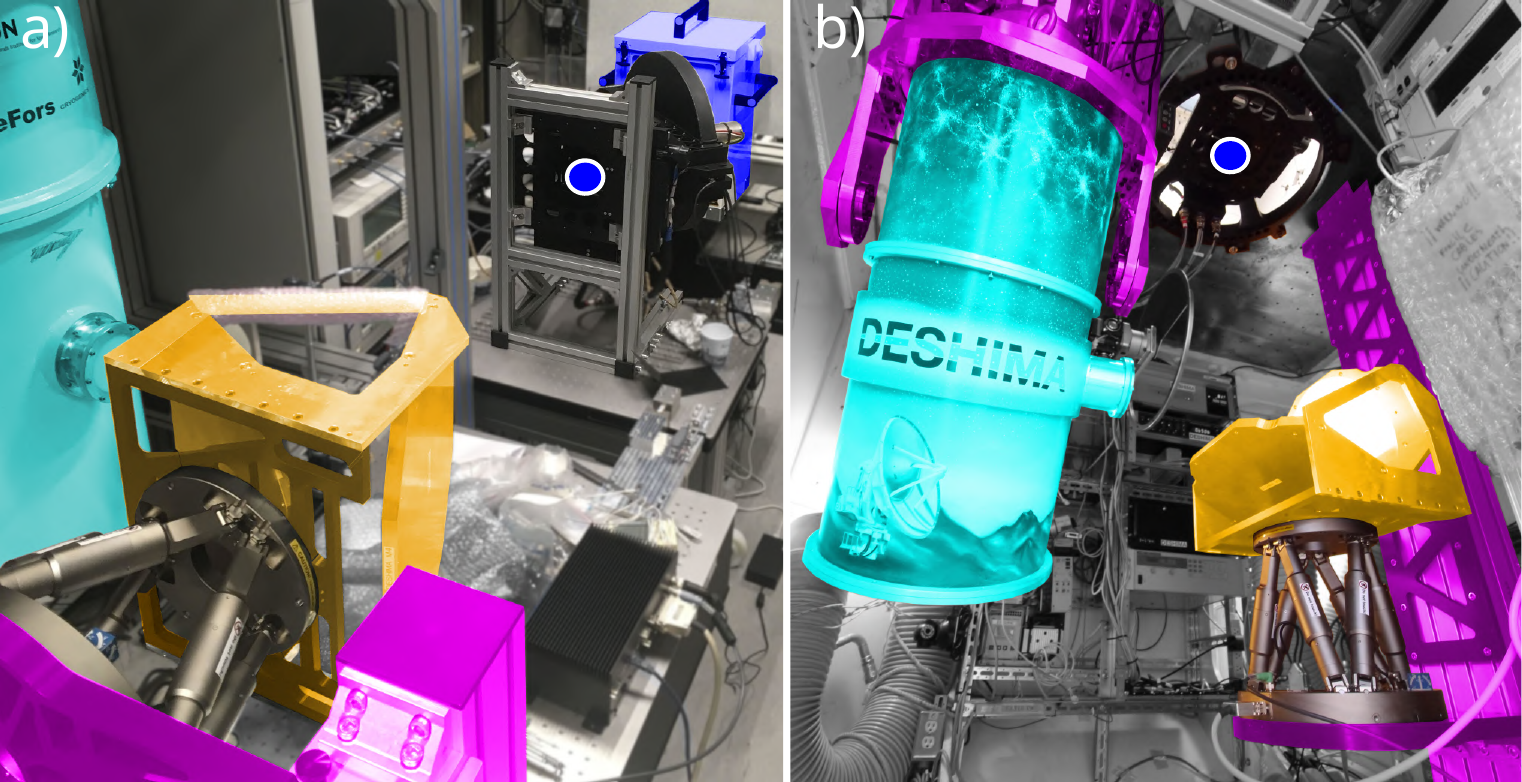}
    \caption{An overview of the two experimental setups used in this work. a) The laboratory setup used for initial characterisation of the alignment method. To the left, in light blue, is the cryostat. The warm optics (orange) are attached to the hexapod, which in turn is attached to the support structure (purple). The cold source, an absorber immersed in a vat containing liquid N$_2$ (dark blue), is placed behind the chopper module (black). The chopper itself is supported by a rectangular frame. The blue disk on the chopper indicates the entrance aperture. b) The setup used inside the ASTE lower cabin. The color coding is the same as for a), but because we now use the atmosphere instead of the cold absorber, there is no blue colored source behind the chopper.}
    \label{fig:lab_aste}
\end{figure*}

\subsection{Warm optics alignment}
\label{subsect:genmethod}
The alignment technique relies on the warm/cold load contrast on the MKIDs as the instrument beam is scanned across the sky chopper entrance aperture using the warm optics mounted on the hexapod. Both paths of the chopper (see Fig.~\ref{fig:chopper}) essentially represent a cylinder, each with a unique exit aperture $A$ and $B$, and a shared entrance aperture. The sky chopper interior and apertures have a physical temperature $T_0$ which we assume to be constant throughout the alignment. 

The atmosphere has an effective physical temperature $T_\mathrm{atm}$, which can be converted into the line-of-sight (LOS) brightness temperature $T_\mathrm{sky}$ using: 
\begin{equation}
\label{eq:Tsky}
    T_\mathrm{sky}(\nu,\mathrm{PWV}) = \left(1 - \eta^{\csc(\epsilon)}_\mathrm{atm}(\nu,\mathrm{PWV}) \right) T_\mathrm{atm},
\end{equation}

where $\eta_\mathrm{atm}(\nu,\mathrm{PWV})$ is the atmospheric transmission coefficient towards zenith, $\nu$ is the frequency at which the atmosphere is observed, PWV is the precipitable water vapor towards zenith, and $\epsilon$ is the elevation of the telescope at which the atmosphere is observed.
We assume for the current treatment that there are only two relevant brightness temperatures, $T_0$ for the sky chopper interior, apertures, and the ASTE receiver cabin in general, and $T_\mathrm{sky}$ for the atmosphere.

Let $\pmb{p}_\mathrm{hex}$ and $\pmb{\theta}_\mathrm{hex}$ denote the hexapod position and orientation, respectively. We define them as follows:
\begin{subequations}
\label{eq:conf}
\begin{align}
    \pmb{p}_\mathrm{hex} &= X \pmb{\hat{x}} + Y \pmb{\hat{y}} + Z \pmb{\hat{z}}, \label{eq:confa}\\
    \pmb{\theta}_\mathrm{hex} &= U \pmb{\hat{\theta}_x} + V \pmb{\hat{\theta}_y} + W \pmb{\hat{\theta}_z}.\label{eq:confb}
\end{align}
\end{subequations}
Note that Eqs.~\ref{eq:confa} and~\ref{eq:confb} represent translations and rotations performed solely by motion of the hexapod and not by how it is mounted or placed. The coordinate system used in Eqs.~\ref{eq:confa} and~\ref{eq:confb} is the same as in Fig.~\ref{fig:sketch_setup}b, with the origin placed in the pivot $\pmb{p}_p$.
We define the hexapod configuration to be the pair $(\pmb{p}_\mathrm{hex},\pmb{\theta}_\mathrm{hex})$ and the hexapod home position to be the configuration $(\pmb{0}, \pmb{0})$, i.e., $X=Y=Z=0$ mm and $U=V=W=0^\circ$. The home configuration is the configuration of the hexapod after installation in the ASTE receiver cabin, before the alignment procedure, and hence has an arbitrary alignment. It therefore serves as an important reference point for comparison of the instrument performance after the alignment procedure.

Let $\delta x(\pmb{p}_\mathrm{hex},\pmb{\theta}_\mathrm{hex})$ denote the response of a single MKID on DESHIMA 2.0 as a function of hexapod configuration. We adopt the definition given in Takekoshi et al.~\cite{Takekoshi2020}:
\begin{equation}
    \delta x(\pmb{p}_\mathrm{hex},\pmb{\theta}_\mathrm{hex}) = \frac{f_0 - f(T_b)}{f_0},
\end{equation}
where $f(T_b)$ represents the resonance frequency of the MKID, loaded at brightness temperature $T_b(\pmb{p}_\mathrm{hex},\pmb{\theta}_\mathrm{hex})$, and $f_0$ the resonance frequency of the MKID, loaded at some reference brightness temperature $T_b = T_0$, which we take as room temperature~\cite{Endo_2019B}. We explicitly include the dependency of $T_b$ on the hexapod configuration to stress the fact that, as the hexapod changes configuration, the beam will be scanned across the sky chopper entrance aperture and hence see a varying brightness temperature distribution. 

Consider the case where the optical system is so misaligned, that the entire beam terminates on the entrance aperture or on the chopper interior. In this case, $T_b \sim T_0$ and $\delta x \sim 0$. When $T_\mathrm{sky} < T_b < T_0$, the beam partly goes through the chopper module and couples to the atmosphere but also partly couples to the chopper aperture/interior, and $\delta x < 0$. In this case, the beam is less misaligned, but the MKID is still partly loaded by the sky chopper aperture/interior. When $T_b \sim T_\mathrm{sky}$, the entire beam couples to the atmosphere. Here, $\delta x$ reaches a minimum and we can conclude that the beam is well-aligned, because it goes through the chopper completely. The goal of the warm optics alignment then is to find a configuration $(\pmb{p}_\mathrm{hex},\pmb{\theta}_\mathrm{hex})$ such that $\delta x$ is minimised.

The alignment can be done with the chopper in either position. This is because the beam from DESHIMA 2.0 scans the entrance aperture, which is shared between beam $A$ and $B$. Also, the distances between the entrance aperture and exit apertures $A$ and $B$ are roughly the same and therefore a misaligned beam, entering through the entrance aperture, will experience the same warm spill-over in both paths. Most important is the presence of two small co-aligned apertures. With only a single aperture, the beam position can be constrained but the beam tilt can still be nonzero. Only upon including the second aperture can the beam tilt also be constrained. 

\subsection{Subreflector alignment}
\label{subsect:subrefalign}
Because the warm optics cannot correct for defocus and lateral misalignment simultaneously~\cite{Moerman2024}, the subreflector is used for refocusing the optical system after the warm optics alignment. The subreflector alignment consists of two steps, which must be repeated at different elevations because of the elevation-dependent gravitational deformation of ASTE~\cite{vonHoerner1975}. The first step consists of scanning the telescope pointing in azimuth and elevation across a small, bright source to measure the peak brightness, and repeating this at various subreflector ($x$, $y$) combinations in order to overlap the Cassegrain focus with the warm focus in the $x$-$y$ plane. The second step consists of scanning the subreflector along the $z$-axis while observing a small, bright source, in order to overlap the Cassegrain focus with the warm focus along the $z$-axis. When the foci overlap, the response of the bright source will be maximal, and the $x$, $y$, and $z$ values corresponding to the maximum is selected as the optimal subreflector ($x$, $y$, $z$) position. The focal optimisation of the subreflector is done using DESHIMA channels at around 350 GHz.

\subsection{Experimental setup}
\label{subsect:expsetup}
We use two different experimental setups for the characterisation, a laboratory and telescope setup. The laboratory setup is designed in such a way to emulate the telescope setup, which can be seen in Fig.~\ref{fig:sketch_setup}b. In this way, we can develop and characterise the alignment method in the laboratory before going to the ASTE telescope to apply the method. We omitted the 30$^\circ$ rotation of the cryostat around the $x$-axis and added a $-$90$^\circ$ rotation of the hexapod, support structure, and warm optics as the beam measurement setup is too large to be placed upright in the laboratory, similar to Moerman et al.~\cite{Moerman2024}. For the definition of the hexapod configuration in the laboratory setup, we also rotate the coordinate system used in Eqs.~\ref{eq:confa} and~\ref{eq:confb} by $-$90$^\circ$ around the $x$-axis, so that it is easier to compare the hexapod characterisation in the laboratory with the characterisation at ASTE. Because it is infeasible to use the atmosphere as the cold source in the laboratory setup we place a cold absorber, immersed in a vat containing liquid nitrogen (N$_2$), behind the sky chopper. Because of the temperature difference between N$_2$ (77 K) and room temperature ($\sim$300 K), we expect ample contrast between the absorber and sky chopper, and hence an accurate alignment. For a photograph of the laboratory setup, see Fig.~\ref{fig:lab_aste}a. In Fig.~\ref{fig:lab_aste}b we show the telescope setup.

\section{Results}
\subsection{Laboratory characterisation}
\label{subsect:labres}
Here we describe the laboratory results of scanning the DESHIMA 2.0 beam across the sky chopper entrance aperture using the hexapod and warm optics with the liquid N$_2$ behind the chopper entrance aperture. We use the DESHIMA channel at 250 GHz for alignment.

\begin{figure}[t]
\centering
    \includegraphics[width=0.5\textwidth]{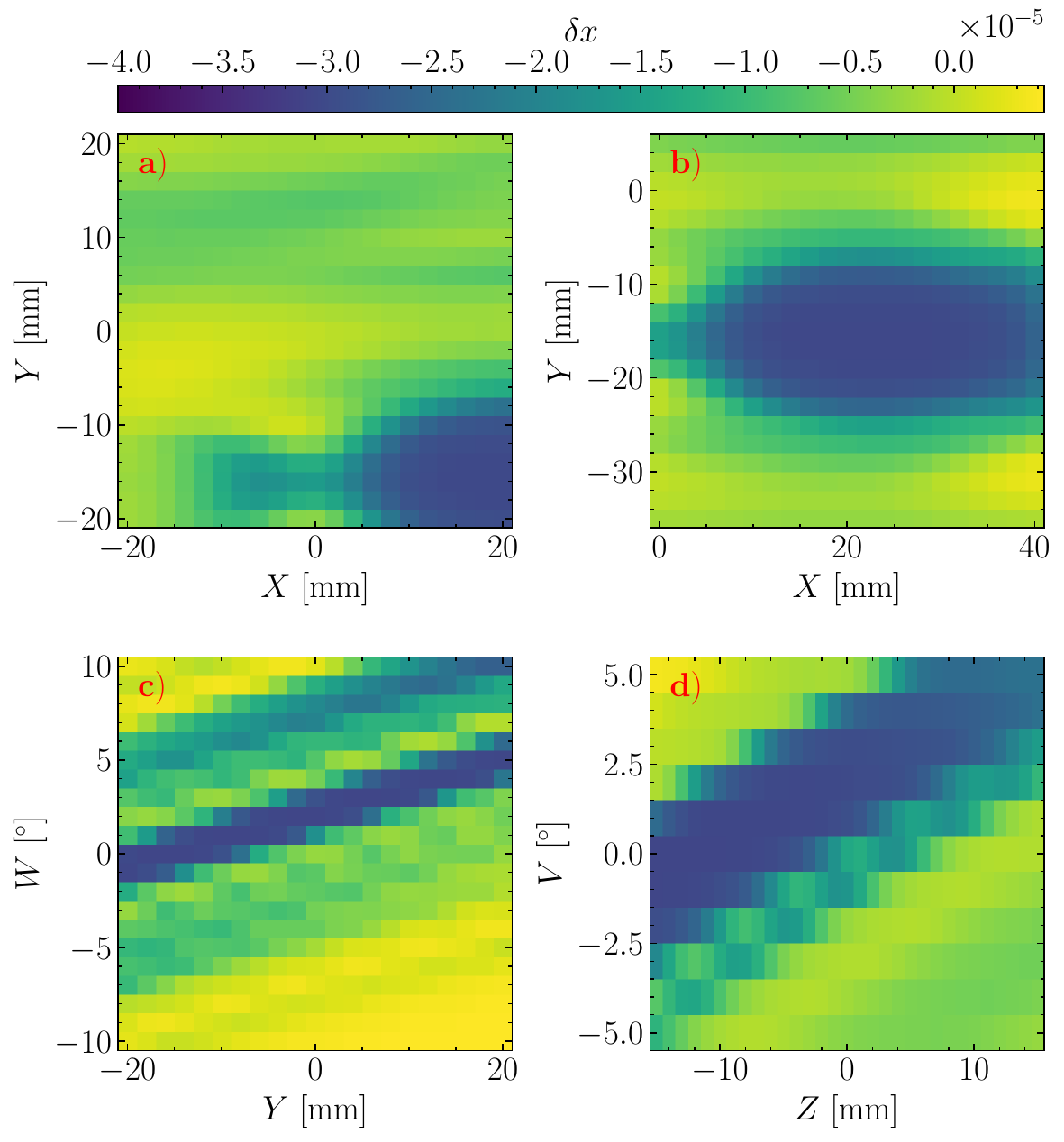}
    \caption{Instrument responses $\delta x$ as function of hexapod scanning DoFs, for the four hexapod scans used for the laboratory characterisation. Shown responses are at a channel frequency $\nu=250$ GHz. We extended the colorbar to $\delta x = -4\cdot10^{-5}$ in order to enhance contrast inside the cold spots.}
    \label{fig:coldspots_lab}
\end{figure}

We started by scanning along the $X$ and $Y$ degrees-of-freedom (DoFs) of the hexapod, centered at $X=Y=0$ mm. We chose this combination because these are non-degenerate, meaning that a change in $X$ or in $Y$ tilts/displaces the beam around/along separate axes, as was shown by the hexapod characterisation performed by Moerman et al.~\cite{Moerman2024}. The result of this initial scan can be found in Fig.~\ref{fig:coldspots_lab}a. We see that $\delta x$ varies significantly over the scan and it appears that the minimum of $\delta x$ is located in the lower right of the scan. Because the DoFs are non-degenerate, we see that there is a single $X$-$Y$ combination for which $\delta x$ reaches a minimum. Smoothly varying either $X$ or $Y$ around the minimum $X$-$Y$ combination smoothly increases $\delta x$, giving rise to a localised $\delta x$ dip, which we call a cold spot. 

For the second $X$-$Y$ scan, we moved the hexapod to $X=20$ mm and $Y=-15$ mm, as suggested by Fig.~\ref{fig:coldspots_lab}a, and scan the DoFs around these values. This result can be found in Fig.~\ref{fig:coldspots_lab}b. It is immediately visible that the minimum of $\delta x$ is now centered on the scan, indicating that this combination of $X$ and $Y$ already produces a better alignment compared to $X=Y=0$ mm. 

After the second $X$-$Y$ scan, we were interested whether we can use $Z$ in the alignment, because the hexapod DoF characterisation performed by Moerman et al.~\cite{Moerman2024} showed that $X$ and $Z$ are degenerate. We did a single $Z$ sweep for the hexapod, setting $X=Y=0$. We found that $Z=-12$ gives the lowest $\delta x$ for this $X-Y$ combination. Because we have no a priori knowledge on whether it is best to use $X$ or $Z$, we pick the DoF that needs the smallest displacement with respect to the home configuration, which is $Z$ in this case.

For the third scan, we scan the $Y$ and $W$ DoFs, setting $Z=-12$ mm and $X=0$ mm. We center $Y$ at 0 mm and $W$ at 0$^\circ$. The results are shown in Fig.~\ref{fig:coldspots_lab}c. These DoFs are degenerate, therefore we do not expect to see a single $Y$-$W$ combination producing the smallest response. Rather, we expect a range of $Y$-$W$ combinations to produce a small $\delta x$, creating a cold valley of minimal $\delta x$ values. We set $W=2^\circ$ and $Y=0$ mm in order to see how a rotational degree of freedom performs. However, from Fig.~\ref{fig:coldspots_lab}c it is clear that any $Y$-$W$ combination lying inside the cold valley is a valid configuration. Had we picked $Y=-15$, $W$ would have been set to 0$^\circ$. 

For the final scan, we verify whether $X=0$ and $Z=-12$ is indeed a good choice. We do this by scanning $Z$ and $V$. In this way, we see if $X=0$ and $Z=-12$ gives the lowest $\delta x$ for $V=0$. If so, we can be sure that the $X$-$Z$ combination is good. The result of which is illustrated in Fig.~\ref{fig:coldspots_lab}d.

From the characterisation, we conclude that a good prescription is to first scan two non-degenerate DoFs, to see where the cold spot is located. We chose $X$ and $Y$, but we showed that one can also change from $X$ to $Z$ as the scanning variable as these two are degenerate. Then, we perform two scans: for each non-degenerate DoF, we perform a scan with that DoF and a DoF that is degenerate. In this way we can fine-tune two degenerate DoFs and ensure that the beam is aligned in the axis affected by the two DoFs.

\begin{figure}[t]
\centering
    \includegraphics[width=0.5\textwidth]{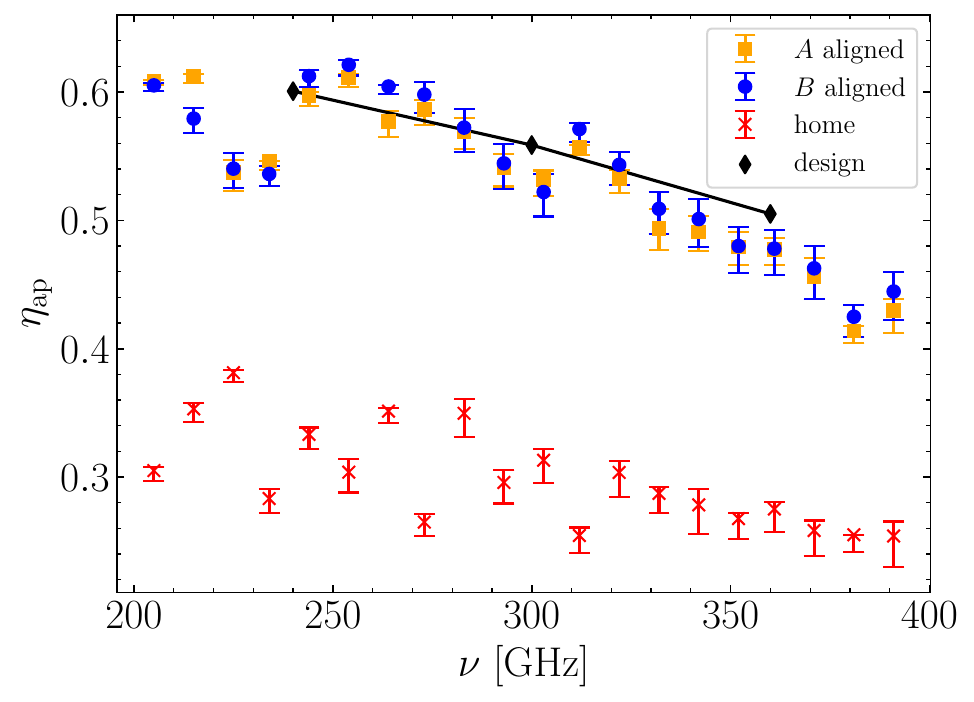}
    \caption{Calculated $\eta_\mathrm{ap}$ as function of channel frequency $\nu$ in GHz for the laboratory characterisation. Red crosses represent the hexapod in home position. The orange boxes and blue dots represent the aligned hexapod position for beams $A$ and $B$, respectively. Black diamonds represent design values. All $\eta_\mathrm{ap}$ are multiplied by the Ruze efficiency $\eta_\rho$. The errors were calculated by propagating the uncertainties in the focus location from Gaussian fitting.}
    \label{fig:etaap_lab}
\end{figure}

After finding the hexapod configuration that corresponded to the cold spot, we measured the complex beam patterns in order to assess the improvement in performance due to the alignment. Using the harmonic phase and amplitude measurement technique described by Moerman et al.~\cite{Moerman2024}, we obtained the complex beam patterns of the home setup and aligned setup, for both beams $A$ and $B$, across the DESHIMA 2.0 bandwidth. By fitting a Gaussian beam to the measured beam patterns, we obtained the tilt and focus location of the Gaussian beam~\cite{Tong2003,Davis_2016}. Using the optical simulation software \texttt{PyPO}~\cite{Moerman2023}, we propagated the measured beam patterns through a model of the ASTE telescope by placing the fitted focus in the cabin focus of the Cassegrain setup at ASTE and calculated the spill-over efficiency $\eta_\mathrm{s}$ on $M_2$, taper efficiency $\eta_\mathrm{t}$ on the aperture subtended by $M_1$, and the aperture efficiency $\eta_\mathrm{ap}\approx\eta_\mathrm{s}\eta_\mathrm{t}$. Because we want to compare the laboratory values to the on-site values, we multiplied $\eta_\mathrm{ap}$ by the Ruze efficiency~\cite{Ruze1966} $\eta_\rho$ of the ASTE telescope to take into account surface roughness, which is given by:
\begin{equation}
    \eta_\rho = \mathrm{exp}\left(-\left( \frac{4\uppi\rho}{\lambda}\right)^2\right),
\end{equation}
where $\rho$ is the root-mean-square (RMS) surface roughness of the ASTE reflectors, which is estimated to be 42 µm, and $\lambda$ the wavelength of incoming radiation. We did not take into account aperture blockage of the primary aperture due to the support struts for $M_5$, as this was not taken into account for the design values. We compared the aligned $\eta_\mathrm{ap}$ to design values. Also, we calculated the $\eta_\mathrm{ap}$ of the hexapod in home configuration. For this calculation of $\eta_\mathrm{ap}$ we used a complex beam pattern measured after removing the entire sky chopper, as the beam was too misaligned and did not pass through the chopper. These results can be found in Fig.~\ref{fig:etaap_lab}.

It is evident from Fig.~\ref{fig:etaap_lab} that the alignment works successfully in the laboratory. The $\eta_\mathrm{ap}$ are at design level, and an average increase of a factor 1.80 for beam $A$ and 1.81 for beam $B$, both with respect to the misaligned case.

\subsection{Telescope results}
\label{subsect:telres}
We report on the alignment for DESHIMA 2.0 at ASTE that was performed on the 7th of November 2023, between 01:15 and 04:50 UTC, fully remotely from the ASTE base camp in San Pedro de Atacama. The telescope elevation during the alignment was 60$^\circ$, which is in the center of the elevation range of ASTE (30$^\circ$--90$^\circ$). We used beam $A$ to align DESHIMA 2.0. Again, we use the DESHIMA channel at 250 GHz for alignment. Using the Atacama Pathfinder EXperiment (APEX)~\cite{Gsten2006} radiometer data we find that the mean PWV during the alignment was about 5.1 mm. At 250 GHz, this corresponds to a $T_\mathrm{sky}$ of about 70 K. Combined with the telescope interior temperature, which is around 293 K, this gives a temperature contrast comparable to the temperature contrast during the laboratory characterisation. Hence, we expect ample contrast for alignment at the telescope. The standard deviation on the mean PWV over the entire alignment was around 0.13 mm, which had a negligible effect on the $T_\mathrm{sky}$ variation at 250 GHz during the alignment. We also note that the PWV during the alignment was high compared to median values recorded for the Chajnantor plateau~\cite{Cortes2020}, where ASTE is situated. Since $T_\mathrm{sky}$ increases monotonically with PWV, a lower PWV enhances the temperature contrast and hence improves the alignment quality.

\begin{figure}[t]
\centering
    \includegraphics[width=0.5\textwidth]{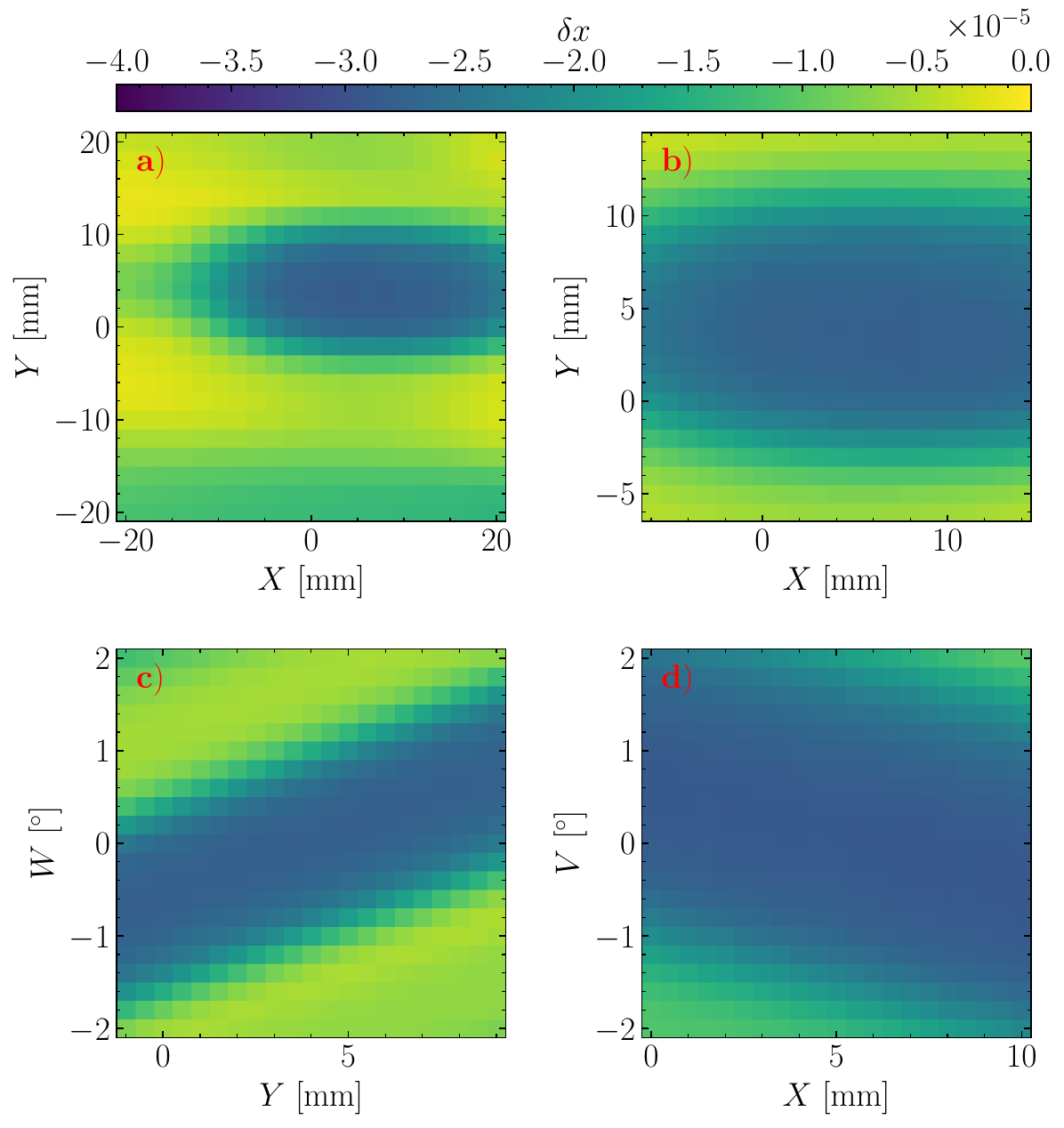}
    \caption{Instrument responses $\delta x$ as function of hexapod scanning DoFs, for the four hexapod scans used for the measurements at ASTE. Shown responses are at a channel frequency $\nu=250$ GHz. The scans were taken at a telescope elevation of 60$^\circ$. Again, we extended the colorbar to $\delta x = -4\cdot10^{-5}$ in order to enhance contrast inside the cold spots.}
    \label{fig:coldspots_tel}
\end{figure}

We first repeated the series of scans taken in Sect.~\ref{subsect:labres} in order to verify the usability of this method at ASTE. The results can be found in Fig.~\ref{fig:coldspots_tel}.  
We have repeated the first two $X$-$Y$ scans to localise the cold spots, as per Sect.~\ref{subsect:labres}. We did not perform a scan along the $Z$ DoF. The $Y$-$W$ scan was taken with $X=5$ mm and centered at $Y=4$ mm. The final $X$-$V$ scan was to check if the previously found parameters indeed create a minimum $\delta x$ value. The difference in $\delta x$ contrast between Fig.~\ref{fig:coldspots_lab} and Fig.~\ref{fig:coldspots_tel} is due to the difference in temperatures of the warm surroundings and cold source in the laboratory and telescope setups.

\begin{figure}[t]
\centering
    \includegraphics[width=0.5\textwidth]{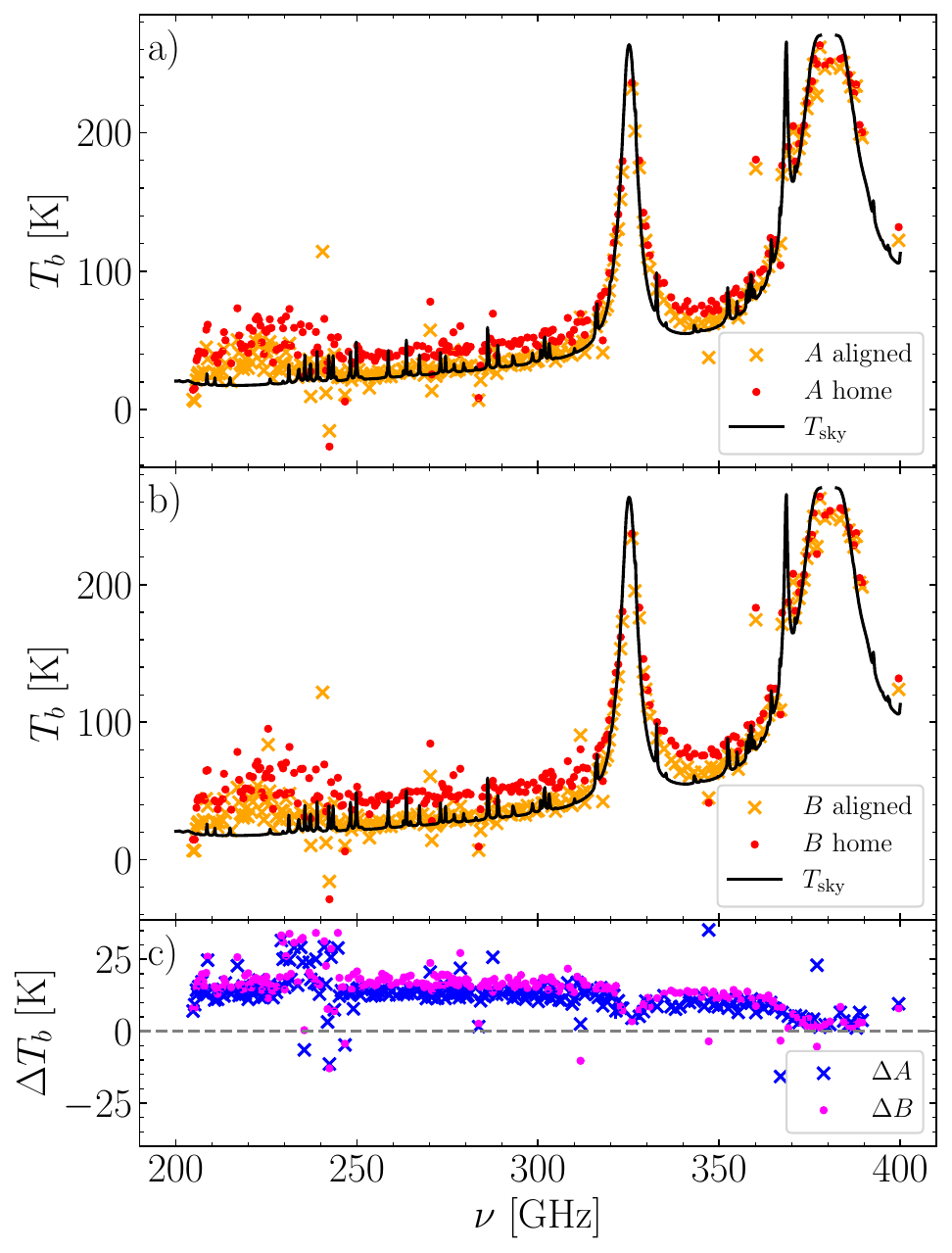}
    \caption{Observations of the atmosphere with beams $A$ and $B$, for both the home and aligned hexapod position. a) Measured $T_b$ in beam $A$ as function of filter frequency $\nu$ for both the home and aligned hexapod configuration. The black line is the theoretical atmosphere $T_\mathrm{sky}$ calculated using ATM~\cite{Pardo2001}, assuming a PWV of 1.22 mm and an elevation of 60$^\circ$. b) same as a), but then for beam $B$. c) The differences between $T_b$ measured in the home and aligned position, for both beam $A$ and $B$. For reference, the grey, horizontal dashed line is plotted at $\Delta T_b = 0$ K.}
    \label{fig:Tsky}
\end{figure}

To verify that the ISS beam is indeed coupled to the atmosphere, we performed four observations of the atmosphere without moving the telescope. Two observations are taken with the hexapod in home position and two with the hexapod in aligned position. For each hexapod position, one observation is performed using beam $A$, the other using beam $B$. We use the skydip method described by Takekoshi et al.~\cite{Takekoshi2020} to convert $\delta x$ to brightness temperature $T_b$.
From the APEX radiometer data we obtained the average PWV during these observations, which turned out to be 1.22 mm, which we then used to calculate $\eta_\mathrm{atm}$ using the Atmospheric Transmission at Microwaves (ATM)~\cite{Pardo2001} software. Then, we calculated the theoretical sky brightness temperature during each observation using Eq.~\ref{eq:Tsky}. A comparison between the measured brightness temperatures $T_b$ in the home and aligned hexapod positions and calculated $T_\mathrm{sky}$ is shown in Fig.~\ref{fig:Tsky}.

From Fig.~\ref{fig:Tsky} we can see that the measured $T_b$ with the aligned hexapod are consistent with the calculated $T_\mathrm{sky}$, indicating that the instrument beam couples to the atmosphere but also that the calibration method described in Takekoshi et al.~\cite{Takekoshi2020} is applicable for a wide-band spectrometer. We find that for lower frequencies, below $\sim$325 GHz, the alignment lowers $T_b$ for both beams $A$ and $B$ by about 20 K compared to the home position, which indicates that the alignment results in less warm spillover and hence a lower background level. Around 325 GHz, there is a strong water emission line which dominates $T_b$ for both the aligned and home positions. Between 325 and 370 GHz the alignment again shows a lower $T_b$, indicating that also at these frequencies, the alignment decreases the warm spillover. Above this frequency, the difference in $T_b$ approaches zero again, as around 370 GHz another strong water emission line is present.

\begin{figure}[t]
\centering
    \includegraphics[width=1\textwidth]{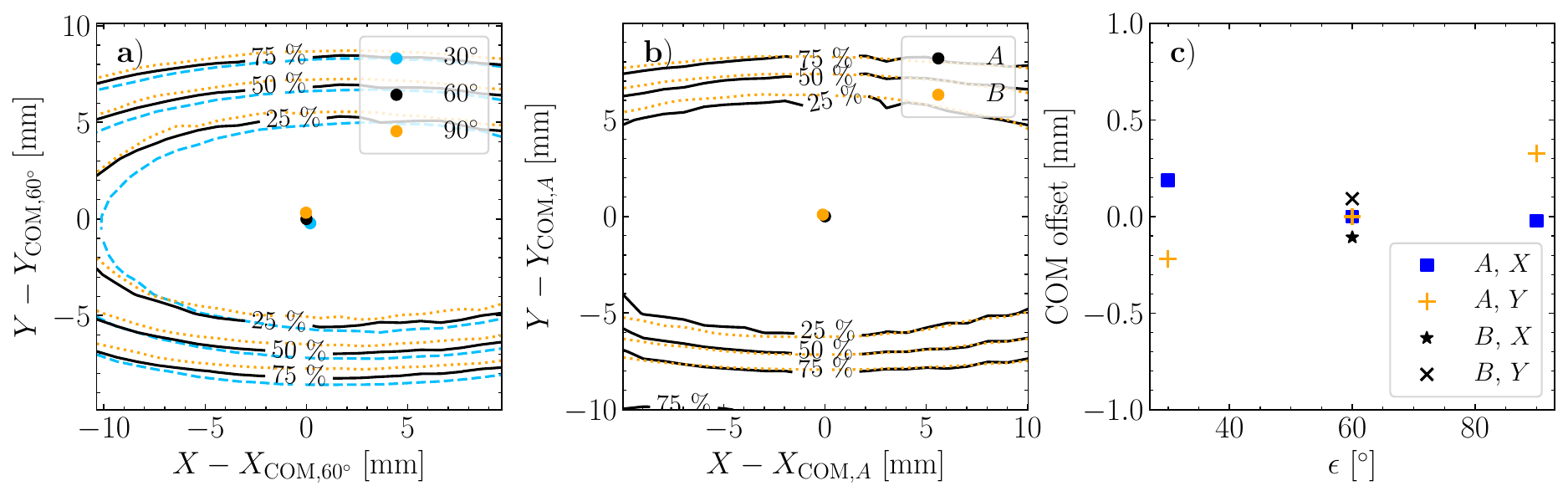}
    \caption{Several hexapod $X$-$Y$ scans to assess alignment stability for different telescope elevations and sky chopper beams. Each scan is normalised such that the minimum $\delta x$ is at 0 and the maximum $\delta x$ at 1. Contours for each scan are shown at 25, 50, and 75\% levels. a) $X$-$Y$ hexapod scan contours and center of mass (COM) for three different telescope elevation angles. The COM positions are denoted by the dots. The $X$ and $Y$ coordinates of the scans are shifted to the COM of the scan at 60$^\circ$. b) $X$-$Y$ hexapod scan contours and COM for beam $A$ and $B$, both at an elevation of 60$^\circ$. In order to faithfully compare beam $A$ and $B$, the scans were taken shortly after each other and therefore the beam $A$ scan at an elevation of 60$^\circ$ in a) is different from the one here, explaining the slightly different cold spot shapes. The $X$ and $Y$ coordinates of the scans are shifted to the COM of the scan taken in beam $A$. c) COM offsets along the $X$ and $Y$ coordinates with respect to the COM of the scan in beam $A$ at elevation angle 60$^\circ$.}
    \label{fig:COMs}
\end{figure}

As a final stability check, we have also performed the $X$-$Y$ scan at an elevation of 90$^\circ$ and 30$^\circ$ and found the same shape of the scan as seen in Fig.~\ref{fig:coldspots_tel}b. We have also calculated the center of mass (COM) for each scan by first normalising the scan so that the maximum $\delta x$ is set to 1 and the minimum to 0. Then, we changed the sign of the normalised scan and added 1, so that the maximum $\delta x$ is at 0 and the minimum at 1. From there, we calculated the COM and found that the COM of the cold spot does not significantly move for different elevations. This indicates that the alignment is stable over elevation angle and could be performed at any elevation angle allowed by the telescope housing the instrument. Also, we have performed the $X$-$Y$ scan using beam $B$ at an elevation of 60$^\circ$. We found again a similar shape to a beam $A$ scan at an elevation of 60$^\circ$ and no significant movement of the COM. The results for the different elevation scans and the beam $B$ comparison can be found in Fig.~\ref{fig:COMs}.

\subsection{Measured aperture efficiency}
\label{subsect:telapereff}
\begin{figure}[t]
\centering
    \includegraphics[width=0.85\textwidth]{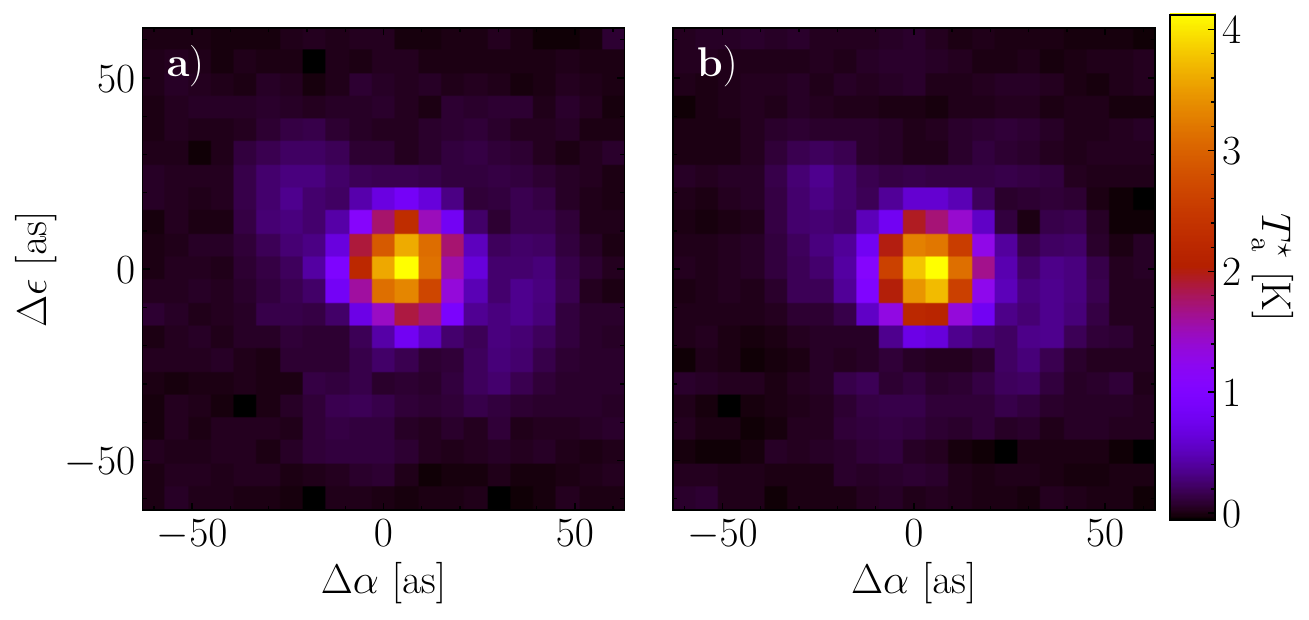}
    \caption{Beam maps for beam $A$ and $B$ at 300 GHz taken on Mars. The azimuth and elevation offsets for both panels are with respect to the center of the scan. a) Beam map taken in beam $A$. b) Beam map taken in beam $B$.}
    \label{fig:maps}
\end{figure}

To calculate the aperture efficiency $\eta_\mathrm{ap}$, we took raster scans of Mars with ASTE, remotely from San Pedro de Atacama using the COSMOS3~\cite{Kamazaki2005} telescope control software. Separate scans were taken for both beam $A$ and $B$. The scans were taken on 17th July 2024. We determined the Martian apparent angular diameter using NASA JPL’s On-Line Solar System Data Service~\cite{Giorgini1996} and adopted $\theta_\mathrm{Mars}=5.62$ arcseconds. For visualisation, the scans at $\nu=300$ GHz can be found in Fig.~\ref{fig:maps}.

Visually, the two scans look very similar, indicating that the alignment has achieved similar quality for beam $A$ and $B$.

\begin{figure}[t]
\centering
    \includegraphics[width=0.5\textwidth]{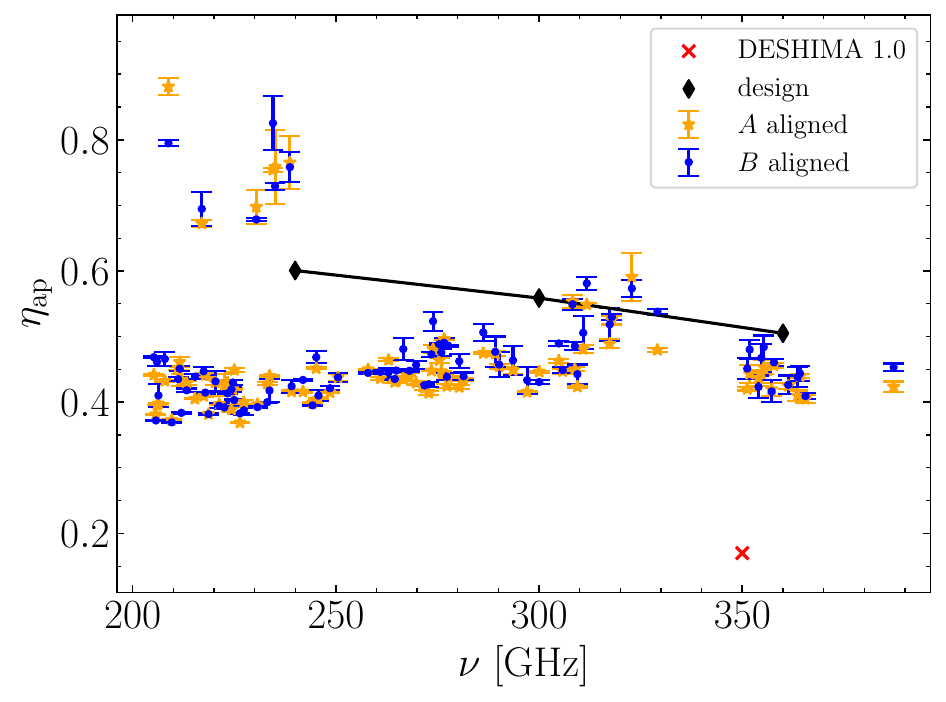}
    \caption{Measured $\eta_\mathrm{ap}$ as function of frequency, obtained from observations of Mars. The red cross denotes the $\eta_\mathrm{ap}$ measured during the DESHIMA 1.0 campaign in 2017, which was 0.17 at 350 GHz. The error bars denote the variance in the measured $\eta_\mathrm{ap}$. We only show $\eta_\mathrm{ap}$ for which the signal-to-noise ratio is larger than 3.}
    \label{fig:etaap_meas}
\end{figure}

We then used the `Mars brightness model' by Lellouch and Amri~\cite{LESIA} to calculate the Martian disk-averaged brightness temperature as function of frequency. Together with the raster scans, $\eta_\mathrm{ap}$ was calculated for both beam $A$ and $B$. The results can be found in Fig.~\ref{fig:etaap_meas}. We calculated the error in $\eta_\mathrm{ap}$, $\sigma_\mathrm{ap}$ from the errors of the fitted parameters in the fitting procedure. In Fig.~\ref{fig:etaap_meas} we only show values for which $\eta_\mathrm{ap} / \sigma_\mathrm{ap} > 3$.

From Fig.~\ref{fig:etaap_meas} it is visible that the measured $\eta_\mathrm{ap}$ for beam $A$ and $B$ are consistent, indicating that the alignment works for both beams. Also, the measured efficiency is larger than the measured $\eta_\mathrm{ap}$ for DESHIMA 1.0 during the 2017 campaign, which was 0.17 at 350 GHz~\cite{Endo_2019B}. By using the alignment procedure described in this work, we have increased $\eta_\mathrm{ap}$ by a factor of 2.5 at 350 GHz. It should be noted that ASTE has not received any upgrades or improvements to the reflector surfaces between the DESHIMA 1.0 and DESHIMA 2.0 campaigns, implying that the surface RMS has not changed. The main difference between the optical systems of DESHIMA 1.0 and DESHIMA 2.0 at ASTE, excluding the sky-position chopper, is the upgrade from a double-slot antenna~\cite{Filipovic1993} to a leaky-lens antenna~\cite{Taniguchi_2022} and a modification of the cold optics to accommodate the change of antenna~\cite{Dabironezare}. This change increased the bandwidth of DESHIMA 2.0 compared to DESHIMA 1.0 but did not change the illumination pattern and edge taper on the subreflector, and hence $\eta_\mathrm{ap}$, of DESHIMA 2.0 compared to DESHIMA 1.0 at 350 GHz.

Most of the measured $\eta_\mathrm{ap}$ are consistently lower than the laboratory values, especially for lower frequencies. The cause of this is a lower than expected $T^\star_\mathrm{a}$, which we suspect is due to an increased sidelobe level. The laboratory values were calculated without accounting for the subreflector support struts, which are known to cause an elevated sidelobe level~\cite{WaiKo1984,Kildal1988}. Error beams arising from the surface roughness of ASTE could also lower the aperture efficiency. They have been proven to be a significant source for pick-up of the IRAM telescope~\cite{Bensch2001}, and could also be significant for ASTE. Methods that take error beams into account exist~\cite{Bensch2001b}, but were not considered given the scope of this work. Another cause for elevated sidelobes could be a slightly elevated edge taper on the ASTE secondary. 

\begin{figure}[t]
\centering
    \includegraphics[width=0.4\textwidth]{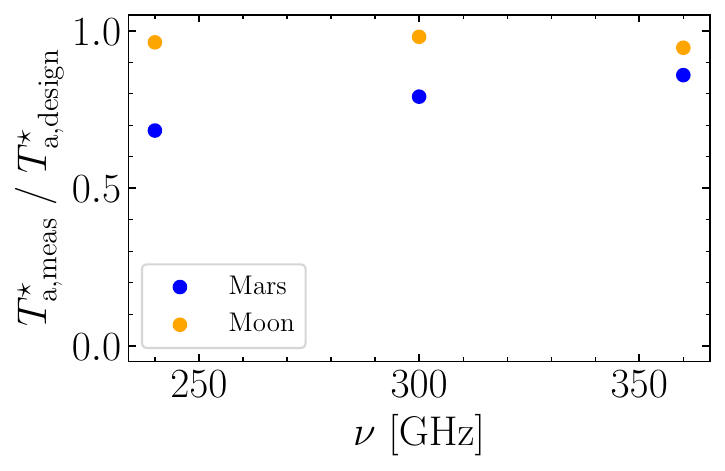}
    \caption{Comparison of the measured $T^\star_\mathrm{a,meas}$ to the expected $T^\star_\mathrm{a,design}$ for both the Moon and Mars beam $A$ observations. For the Moon, we used the brightness temperature model from Mangum~\cite{Mangum1993} and assumed this to be equal to $T^\star_\mathrm{a,design}$. For Mars, we used the brightness temperature model by Lellouch and Amri~\cite{LESIA} and the design $\eta_\mathrm{ap}$ and DESHIMA beamsizes to calculate $T^\star_\mathrm{a,design}$.}
    \label{fig:ratios}
\end{figure}

We confirmed the elevated sidelobe level by observing the Moon, which is of sufficient angular diameter to fill the sidelobes of the DESHIMA beam pattern and any error beams. This observation was compared to a brightness temperature model of the Moon given by Mangum~\cite{Mangum1993} and showed that $T^\star_\mathrm{a}$ at 240, 300, and 360 GHz was recovered to a level within 95\% of the expected value. In Fig.~\ref{fig:ratios} we show this result, together with the beam $A$ observation of Mars which shows the lower than expected $T^\star_\mathrm{a}$.

Conversely, there are six channels between 210--240 GHz that have $\eta_\mathrm{ap}$ values about twice as large as the rest of the $\eta_\mathrm{ap}$ in this range. Upon inspection, we found that the fitted $T^\star_\mathrm{a}$ for these channels were about twice as large as the surrounding channels, explaining the discrepancy. We also checked the central and maximal values of the Mars maps directly and found the same discrepancy, which indicated that the fitting process was not the cause. In addition, we inspected the off-source atmospheric spectrum from the outskirts of the Mars maps and also found elevated $T^\star_\mathrm{a}$ values for these channels. Since the atmosphere is large enough to fill the entire DESHIMA beam at all channel frequencies, these observed $T^\star_\mathrm{a}$ are independent of $\eta_\mathrm{ap}$ and hence could indicate an issue in the responsivity calibration for these particular channels.

\section{Conclusion}
We have developed an alignment method for the DESHIMA 2.0 spectrometer at ASTE, which uses the contrast of the cold atmosphere against the warm interior of the sky chopper together with a modified Dragonian dual reflector and mechanical hexapod, and can be remotely performed over the internet.

Before we applied the alignment method to DESHIMA 2.0 at ASTE, we characterised the alignment method in the laboratory, using the same DESHIMA 2.0 chip as in Chile. We placed an absorber, immersed in a vat containing liquid N$_2$, behind the sky chopper in order to emulate the cold sky. The laboratory measurements show that the alignment method successfully recovers the designed performance, starting from a misaligned system. 

The alignment was then applied to the optical chain of DESHIMA 2.0 at ASTE. We showed that the characterisation and alignment procedure, which we tested in the laboratory, is identical at the telescope, using the atmosphere instead of N$_2$. This proves that the atmosphere is a suitable cold background against the sky chopper interior and can be used for the described alignment method. Additionally, we showed observations of the atmosphere performed with DESHIMA 2.0 at ASTE and compared these to the theoretical atmosphere brightness temperature curve and found that the alignment causes less warm spillover for both beams $A$ and $B$ compared to the hexapod home position. As a final test, we measured the aperture efficiency using Mars, and showed that we have increased the efficiency by a factor 2.5 compared to DESHIMA 1.0 during its 2017 campaign. These observations provide evidence for the successful alignment using our method. However, we find that the measured efficiency did not reach design values and we suspect a raised sidelobe level or error beams to cause this.

The alignment method can in principle be applied to any single-pixel sub-mm spectrometer. In this work we used the sky chopper, which acted as a cylinder with a higher temperature than the atmosphere. The role of the cylinder can be played by any cylinder-like object, provided that its brightness temperature is sufficiently higher than that of the atmosphere.

\section{Appendix: Determination of aperture efficiency from planet map}
\label{appendix:apereff}
In order to obtain the aperture efficiency $\eta_\mathrm{ap}$ from a planet map, a set of parameters needs to be extracted. To obtain these map parameters, a disk-convolved 2D Gaussian fit was performed to each channel map. Throughout this explanation, we will denote the corrected antenna temperature map $T^\star_\mathrm{a}(\alpha,\epsilon)$, where $\alpha$ and $\epsilon$ represent the horizontal (Azimuth) and vertical (Elevation) coordinates, respectively. In our case, we bin the full TOD in Azimuth and Elevation pixels of a size of 6 by 6 arcseconds. Each pixel in $T^\star_\mathrm{a}(\alpha,\epsilon)$ has an associated standard deviation denoted $\sigma_\mathrm{a}(\alpha,\epsilon)$. The conversion of measured antenna temperature $T_\mathrm{a}$ to $T^\star_\mathrm{a}$ was carried out using the chopper wheel calibration method~\cite{Ulich1976}.
As initial parameters for the Gaussian to be fitted, we used the statistical moments of the map as described by Burger and Burge~\cite{Burger2016}. The raw moments $M_{ij}$ and central moments $\mu_{ij}$ are given by:
\begin{align}
    M_{ij} &= \sum_\alpha \sum_\epsilon \alpha^i \epsilon^j T^\star_\mathrm{a}(\alpha,\epsilon),\\
    \mu_{ij} &= \sum_\alpha \sum_\epsilon (\alpha-\bar{\alpha})^i (\epsilon-\bar{\epsilon})^j T^\star_\mathrm{a}(\alpha,\epsilon),
\end{align}
where $i,j \in \mathbb{N}$ and $\bar{\alpha}$ and $\bar{\epsilon}$ are the central $\alpha$ and $\epsilon$ values:
\begin{equation}
    (\bar{\alpha}, \bar{\epsilon}) = \left( \frac{M_{10}}{M_{00}}, \frac{M_{01}}{M_{00}} \right).
\end{equation}
We estimate the floor level $T^\star_\mathrm{a,floor}$ by taking the average of the off-source sections of the $T^\star_\mathrm{a}(\alpha,\epsilon)$ map. The amplitude $T^\star_\mathrm{a,amp}$ is then estimated by taking the maximum of $T^\star_\mathrm{a}(\alpha,\epsilon) - T^\star_\mathrm{a,floor}$.

The position angle $\Psi$ of the 2D Gaussian, with respect to the positive $\alpha$-axis, is given by:
\begin{equation}
    \Psi = \frac{1}{2} \arctan \left( \frac{2\mu_{11}}{\mu_{20} - \mu_{02}} \right),
\end{equation}
and the eccentricity $e$ is given by:
\begin{equation}
    e = \frac{(\mu_{20} - \mu_{02})^2 + 4 \mu^2_{11}}{(\mu_{20}+\mu_{02})^2}.
\end{equation}
For the semi-minor axis $w_y$, we use the diffraction limited full-width-half-maximum (FWHM) for cosine-tapered illumination~\cite{Condon2016}, calculated using:
\begin{equation}
    w_y \approx 1.2\frac{\lambda}{2 R_\mathrm{tel}},
\end{equation}
where $\lambda$ is the wavelength of the radiation at which the map was taken and $R_\mathrm{tel}$ is the radius of the primary aperture of the telescope. For ASTE, $R_\mathrm{tel}=5$ meter. Because we define the semi-major axis $w_x$ to lie along the $x$-axis (without rotation applied), we calculate the initial estimate to be:
\begin{equation}
\label{eq:wx}
    w_x = w_y \left( 1-e^2 \right)^{-1/2}.
\end{equation}

We then generate a Gaussian $\mathcal{G}(\alpha,\epsilon)$ using the following expression:
\begin{align}
\label{eq:Gauss}
    a &= \frac{\cos^2 \Psi}{2\sigma^2_x} + \frac{\sin^2 \Psi}{2\sigma^2_y},\\
    b &= -\frac{\sin\Psi\cos\Psi}{2\sigma^2_x} + \frac{\sin\Psi\cos\Psi}{2\sigma^2_y},\\
    c &= \frac{\sin^2 \Psi}{2\sigma^2_x} + \frac{\cos^2 \Psi}{2\sigma^2_y},\\
    \mathcal{G}(\alpha,\epsilon) &= \exp\left( - \left(a(\alpha-\bar{\alpha})^2 + b(\alpha-\bar{\alpha})(\epsilon-\bar{\epsilon}) + c(\epsilon-\bar{\epsilon})^2 \right) \right).
\end{align}
Note that we use the standard deviation in Eq.~\ref{eq:Gauss} instead of the FWHM. We can convert from some FWHM $w$ to standard deviation $\sigma$ using:
\begin{equation}
    \sigma = \frac{w}{2\sqrt{2\ln 2}}.
\end{equation}

We then define a disk $\mathcal{D}(\alpha,\epsilon)$ with angular diameter $\theta$ equal to the apparent angular diameter of the planet that is scanned. Each point on the disk is assigned a brightness temperature $T_b$, which we take as the disk-averaged brightness temperature for the planet in question:
\begin{equation}
    \mathcal{D}(\alpha,\epsilon) =
    \begin{cases}
        T_b & \text{if } \sqrt{\alpha^2+\epsilon^2} \leq \theta,\\
        0 & \text{otherwise}.
    \end{cases}
\end{equation}

The Gaussian is normalised and convolved with the disk to obtain the disk-convolved Gaussian $\mathcal{GD}(\alpha,\epsilon)$:
\begin{equation}
\label{eq:gd_conv}
    \mathcal{GD}(\alpha,\epsilon) = \frac{\sum_{\alpha'}\sum_{\epsilon'}\mathcal{D}(\alpha',\epsilon')\mathcal{G}(\alpha-\alpha',\epsilon-\epsilon')\Delta \alpha'\Delta \epsilon'}{\sum_{\alpha}\sum_{\epsilon}\mathcal{G}(\alpha,\epsilon) \Delta \alpha\Delta \epsilon}. 
\end{equation}
For the initial estimate for the fit we set $T_b=1$, normalise $\mathcal{GD}(\alpha,\epsilon)$ to unit amplitude, multiply by $T^\star_\mathrm{a,amp}$ and add $T^\star_\mathrm{a,floor}$. For the subsequent fitting procedure however, we do not normalize to unit amplitude and just add $T^\star_\mathrm{a,floor}$ to $\mathcal{GD}(\alpha,\epsilon)$. We weigh each point in $T^\star_\mathrm{a}(\alpha, \epsilon)$ by $\sigma^{-2}_\mathrm{a}(\alpha, \epsilon)$.

Using the \texttt{curve\_fit} function in the \texttt{scipy}~\cite{Virtanen2020} package, we fit $\mathcal{GD}(\alpha,\epsilon)$ to $T^\star_\mathrm{a}(\alpha,\epsilon)$. The free parameters are $T_b, w_y, \bar{\alpha},\bar{\epsilon},T^\star_\mathrm{a,floor},e$, and $\Psi$. After fitting, the relevant parameters for the aperture efficiency calculation are denoted $\hat{T}_b,\hat{e}$ and $\hat{w}_y$, from which we can calculate $\hat{w}_x$ using Eq.~\ref{eq:wx}. With $\hat{T}_b$ and a brightness temperature model $T_{b,\mathrm{model}}$ of the observed planet, we calculate the main beam efficiency:
\begin{equation}
    \eta_\mathrm{mb} = \frac{\hat{T}_b}{T_{b,\mathrm{model}}},
\end{equation}

From $\hat{w}_x$ and $\hat{w}_y$, we calculate the main beam solid angle:
\begin{equation}
    \Omega_\mathrm{mb} = \frac{\uppi \hat{w}_x \hat{w}_y}{4\ln 2},
\end{equation}
and using $\eta_\mathrm{mb}$ and $\Omega_\mathrm{mb}$ we finally calculate the aperture efficiency:
\begin{equation}
    \eta_\mathrm{ap} = \frac{\lambda^2 \eta_\mathrm{mb}}{A_\mathrm{p} \Omega_\mathrm{mb}},
\end{equation}
where $A_\mathrm{p} = \uppi R^2_\mathrm{tel}$ is the physical area of the primary aperture.

\subsection* {Disclosures}
The authors declare that there are no financial interests, commercial affiliations, or other potential conflicts of interest that could have influenced the objectivity of this research or the writing of this paper.

\subsection* {Code, Data, and Materials Availability} 
The processed data and code used for generating the figures in this paper are available on Zenodo: \url{https://zenodo.org/records/15229117}. The raw, unprocessed data is approximately 31.5 gigabytes in size and is not publicly available due to its large size. The raw data are available upon request from the corresponding author.

\subsection* {Acknowledgments}
This work is supported by the European Union (ERC Consolidator Grant No. 101043486 TIFUUN). Views and opinions expressed are however those of the authors only and do not necessarily reflect those of the European Union or the European Research Council Executive Agency. Neither the European Union nor the granting authority can be held responsible for them. TT was supported by the MEXT Leading Initiative for Excellent Young Researchers (Grant No. JPMXS0320200188). This study was supported by JSPS KAKENHI Grant Numbers JP22H04939, JP23K20035, and JP24H00004. The ASTE telescope is operated by National Astronomical Observatory of Japan (NAOJ).


\bibliography{article}   
\bibliographystyle{spiejour}   


\vspace{2ex}\noindent\textbf{Arend Moerman} received the MSc degree in astronomy and astrophysics from Leiden University in 2022. He is currently with Delft University of Technology, where he is pursuing the doctoral degree. His current research interest is focused on galaxy cluster dynamics and instrumentation for the mm/sub-mm  wavelength range.

\vspace{2ex}\noindent\textbf{Kenichi Karatsu} is an Instrument Scientist with SRON. He received his Ph.D. degree in science in 2011 from Kyoto University with study of proton spin structure at the RHIC-PHENIX experiment. In 2015, He joined the DESHIMA project at TU Delft/SRON. His main role is to lead laboratory evaluation and telescope campaign of the instrument. His research interest is to develop an experimental instrument for revealing mysteries of the Universe.

\vspace{2ex}\noindent\textbf{Jochem Baselmans} is full professor in experimental astronomy at the Delft University of Technology, senior instrument scientist at the SRON Netherlands Institute for Space Research and visiting professor at the University of Cologne. He is an expert in the development of instrumentation for far-infrared astronomy based upon Microwave Kinetic Inductance Detectors. He invented the concept of the on-chip spectrometer for far-infrared radiation and received an ERC-COG to develop this technology. He is the lead system engineer for the first on-chip spectrometer instrument, DESHIMA, fielded on the 10m ASTE telescope in 2019 and its follow-up instrument DESHIMA 2.0. He is the detector lead for the new AMKID camera on APEX and has pushed MKID technology to sensitivity levels suitable for future cryogenically cooled space-based observatories.

\vspace{2ex}\noindent\textbf{Shahab Oddin Dabironezare} received the Ph.D. degree in electrical engineering from Delft University of Technology (TU Delft), Delft, The Netherlands, in 2020. He is currently an Assistant Professor in the THz Sensing Group at TU Delft and instrument scientist at the Netherlands Institute for Space Research (SRON). His research interests include wideband antennas at millimeter and submillimeter-wave applications, wide field-of-view imaging systems, quasi-optical systems, lens antennas, and absorber-based passive cameras.

\vspace{2ex}\noindent\textbf{Shinji Fujita} is a project assistant professor at The Institute of Statistical Mathematics, Japan. His research focuses on molecular clouds and star formation in the Milky Way Galaxy, using radio telescopes such as ASTE and the Nobeyama 45m telescope. He employs data science approaches including machine learning and deep learning to analyze large-scale astronomical data.

\vspace{2ex}\noindent\textbf{Robert Huiting} is a mechanical design engineer. He received his BS degree precision engineering (2001) and technology management (2003) at the Hogeschool van Utrecht. He has been working as a design engineer for SRON Netherlands institute for space research since 2007. He has worked as a research engineer at the FOM institute for plasma physics for the XUV optics group the years before. His experience from the last 13 years lies in designing instrumentation for ground based space research.

\vspace{2ex}\noindent\textbf{Kotaro Kohno} is a professor and director at the Institute of Astronomy, the University of Tokyo. He received his PhD in 1998 for studies of dense molecular gas in Seyfert and starburst galaxies. His research focuses on spatially resolved studies of galaxies using ALMA and JWST, and on surveys at submm/mm/radio wavelengths using techniques such as gravitational lensing, radio-optical crossmatching, and line intensity mapping. He is also actively involved in instrumentation development for these studies.

\vspace{2ex}\noindent\textbf{Yuri Nishimura} is a project assistant professor at the Institute of Astronomy, the University of Tokyo, focusing on integrated superconducting spectrometer development and analysis. She received her PhD in physics from the University of Tokyo in 2017. Her research interests include the physics and chemistry of the interstellar medium, particularly on scales ranging from molecular clouds to entire galaxies, in the context of galaxy evolution.

\vspace{2ex}\noindent\textbf{Fenno Steenvoorde} is an instrument maker at DEMO, TU Delft. He graduated from the Leidse instrumentmakers School in 2020, and worked for SRON Netherlands Institute for Space Research during 2020--2022. As a member of the DESHIMA team, he developed cryogenic and optical components, and participated in the installation on the ASTE telescope in 2023. His current interest includes the development of the cryogenic components for TIFUUN.

\vspace{2ex}\noindent\textbf{Tatsuya Takekoshi} is an assistant professor at Kitami Institute of Technology, specializing in submillimeter and millimeter wave instrumentation for astronomy. His research focuses on star formation and observational studies using the ASTE telescope. He is an expert in optical and mechanical design for astronomical instrumentation, contributing to the development of advanced observing systems, including cryogenic and superconducting detector technologies.

\vspace{2ex}\noindent\textbf{Yoichi Tamura} is Professor of Astronomy at Nagoya University. He received his Ph.D. from the University of Tokyo and has worked for National Astronomical Observatory of Japan and the University of Tokyo before joining Nagoya University. He brings a 20-year experience in millimeter and submillimeter astronomy and instrumentation. His main interests include a conceptual design of a future large-aperture submillimeter telescope, commissioning and signal processing of an integrated superconductor spectrometer and a coherent receiver system to existing millimeter and submillimeter telescopes. These things facilitate his astronomical interests in galaxy formation in the early Universe.

\vspace{2ex}\noindent\textbf{Akio Taniguchi} is a project assistant professor at Kitami Institute of Technology. He received his PhD from the University of Tokyo in 2018 and has worked as a postdoc and a project assistant professor at Nagoya University before joining Kitami Institute of Technology. His research interests focus on the development of efficient spectroscopy for next-generation submillimeter single-dish telescopes based on data scientific approaches. He is contributing to the development of data analysis softwares for the submillimeter spectroscopic instruments such as DESHIMA, to the data analysis methods implemented on top of them, and to the commissioning and science verification using them.

\vspace{2ex}\noindent\textbf{Stephen Yates} is an instrument scientist at SRON, the Netherlands Institute for Space Research (since 2006). He received a Ph.D. from the University of Bristol (2003), followed by work at the CNRS-CRTBT (now Institut Neél) Grenoble on experimental low temperature physics (2003--2006). His current interests concentrate on MKID development for sub-mm to FIR astronomical applications but also include a wider interest in device physics and superconductivity, optics, and full end to end instrument characterization and performance. Dr. Yates has published over 80 papers. 

\vspace{2ex}\noindent\textbf{Bernhard R. Brandl} is a professor of infrared astronomy, at Leiden University and Technical University Delft, who works on starburst galaxies, HII regions, and astronomical instrumentation. He has authored over 130 papers in refereed astronomical journals and over 150 technical publications, with more than 15000 citations. He has been involved in the design and construction of the JWST-MIRI spectrometer, and he is the Principal Investigator of the 1st-generation instrument METIS for ESO’s Extremely Large Telescope.

\vspace{2ex}\noindent\textbf{Akira Endo} is an experimental astronomer and associate professor at TU Delft. He is interested in mapping the large-scale Universe in 3D, and the required development of (sub)millimeter wave imaging spectrometers. He leads the DESHIMA collaboration as the Dutch principal investigator, and currently leads an ERC Consolidator grant collaboration to develop TIFUUN: an integral field spectrometer based on plug-and-play, open-hardware integral field units (IFUs) that are individually designed to address specific astronomical questions.



\end{document}